\documentclass[journal]{IEEEtran}
 
\makeatletter
\def\ps@headings{%
	\def\@oddhead{\mbox{}\scriptsize\rightmark \hfil \thepage}%
	\def\@evenhead{\scriptsize\thepage \hfil \leftmark\mbox{}}%
	\def\@oddfoot{}%
	\def\@evenfoot{}}
\makeatother 

\makeatletter
\def\endthebibliography{%
	\def\@noitemerr{\@latex@warning{Empty `thebibliography' environment}}%
	\endlist
}
\makeatother

\usepackage{graphicx}
\usepackage{amsmath, amsfonts,epsfig, multirow, floatflt}
\usepackage{amssymb}
\usepackage{subfigure}
\usepackage{cite}
\usepackage{algorithm}
\usepackage{algorithmicx}
\usepackage{algpseudocode}
\usepackage{enumitem}
\usepackage{diagbox}
\usepackage{mathrsfs}
\usepackage{url}
\usepackage{color}

\usepackage{caption}
\usepackage{hhline}
\usepackage{array}
\usepackage{booktabs}
\usepackage{amsthm}
\usepackage{bm}
\usepackage{setspace}
\usepackage[table,xcdraw]{xcolor}
\usepackage{stfloats}

\newcommand{\MYnewpage}{%
	\ifCLASSOPTIONonecolumn
		\ifCLASSOPTIONjournal
			\typeout{The onecolumn journal mode.}
			\newpage
		\fi
	\fi}

\allowdisplaybreaks[4]

\begin{document}
%% *************************************************************************
\ifCLASSOPTIONonecolumn
    \typeout{The onecolumn mode.}
    \title{VoI-Driven Joint Optimization of Control and Communication 
in Vehicular Digital Twin Network}
   
    \author{Lei~Lei,~\IEEEmembership{Senior~Member,~IEEE}, Kan~Zheng,~\IEEEmembership{Fellow,~IEEE},  Jie~Mei,~\IEEEmembership{~Member,~IEEE}, and~Xuemin (Sherman)~ Shen,~\IEEEmembership{Fellow,~IEEE}% <-this % stops a space
	
 \thanks{Lei Lei is with the School of Engineering, University of Guelph, N1G 2W1, Canada.}% <-this % stops a space 
  \thanks{Kan Zheng  and Jie Mei are with the College of Electrical Engineering and Computer Sciences, Ningbo University, 315211, China.}% <-this % stops a space
        
        \thanks{ Xuemin (Sherman) Shen is with the Department of Electrical and Computer Engineering, University of Waterloo, N2L 3G1, Canada .}% <-this % stops a space

    }
\else
    \typeout{The twocolumn mode.}
    \title{VoI-Driven Joint Optimization of Control and Communication 
in Vehicular Digital Twin Network}
    \author{Lei~Lei,~\IEEEmembership{Senior~Member,~IEEE}, Kan~Zheng,~\IEEEmembership{Fellow,~IEEE}, Jie~Mei,~\IEEEmembership{Member,~IEEE}, \\ and~Xuemin (Sherman)~Shen,~\IEEEmembership{Fellow,~IEEE}% 
  
   \thanks{Lei Lei is with the School of Engineering, University of Guelph, N1G 2W1, Canada.}% <-
  \thanks{Kan Zheng  and Jie Mei are with the College of Electrical Engineering and Computer Sciences, Ningbo University, Ningbo, 315211, China.}% <-this % 
        
        \thanks{ Xuemin (Sherman) Shen is with the Department of Electrical and Computer Engineering, University of Waterloo, N2L 3G1, Canada.}% <-
    }
\fi

\ifCLASSOPTIONonecolumn
	\typeout{The onecolumn mode.}
\else
	\typeout{The twocolumn mode.}
%	\markboth{}{Author \MakeLowercase{\textit{et al.}}: Title}
\fi

\maketitle

\ifCLASSOPTIONonecolumn
	\typeout{The onecolumn mode.}
	\vspace*{-50pt}
\else
	\typeout{The twocolumn mode.}
\fi
\begin{abstract}

The vision of sixth-generation (6G) wireless networks paves the way for the seamless integration of digital twins into vehicular networks, giving rise to a Vehicular Digital Twin Network (VDTN). The large amount of computing resources as well as the massive amount of spatial-temporal data in Digital Twin (DT) domain can be utilized to enhance the communication and control performance of Internet of Vehicle (IoV) systems. In this article, we first propose the architecture of VDTN, emphasizing key modules that center on functions related to the joint optimization of control and communication. We then delve into the intricacies of the multitimescale decision process inherent in joint optimization in VDTN, specifically investigating the dynamic interplay between control and communication. To facilitate the joint optimization, we define two Value of Information (VoI) concepts rooted in control performance. Subsequently, utilizing VoI as a bridge between control and communication, we introduce a novel joint optimization framework, which involves iterative processing of two Deep Reinforcement Learning (DRL) modules corresponding to control and communication to derive the optimal policy. Finally, we conduct simulations of the proposed framework applied to a platoon scenario to demonstrate its effectiveness in ensuring the stability and transmission efficiency of VDTN.

\end{abstract}

\ifCLASSOPTIONonecolumn
	\typeout{The onecolumn mode.}
	\vspace*{-10pt}
\else
	\typeout{The twocolumn mode.}
\fi

\IEEEpeerreviewmaketitle

\MYnewpage

\section{Introduction}
\label{sec:Introduction}

\IEEEPARstart{T}he evolution of Internet of Vehicles (IoV) is closely connected to the ongoing advancements in the fifth generation (5G) Cellular Vehicle-to-Everything (C-V2X) communication technology. Although 5G C-V2X has ushered in impressive improvements in the communication capabilities for IoV, there are still critical challenges in supporting a wide range of safety and non-safety IoV applications~\cite{5GV2X}. The inherent promise of six generation (6G) lies in its ability to deliver significantly increased throughput and reduced communication latency. These advancements are expected to provide the essential infrastructure needed to overcome the current constraints in IoV. The enhanced capabilities of 6G also open doors for effective integration with edge and cloud computing resources, alleviating the burden on individual vehicles~\cite{6GIoV}.

The rapid advancement of Digital Twin (DT) technologies, especially with robust support by 6G, is poised to revolutionize the IoV~\cite{6GDT}. This revolution is exemplified by the emergence of the Vehicular Digital Twin Network (VDTN), characterized by the seamless integration of DT into vehicular networks. In essence, a DT is a real-time evolving digital replica used to emulate a physical object or process, encapsulating its entire history in the virtual space~\cite{6GDT2}. Within the VDTN, comprehensive virtual representations of objects or entities, such as vehicles in physical space, can be created and maintained to establish the DT. As a result, the VDTN has the potential to leverage DT to enhance communication, control, and overall system performance. It stands as a promising candidate for the next generation of IoV, offering new dimensions of efficiency, reliability, and intelligence through the synergistic integration of DT and 6G technologies.

Among the vast applications of IoV, the efficient and reliable control of autonomous vehicles (AVs) is a critical task, which encompasses both the control and communication subsystems. Traditionally, these subsystems are often designed independently, utilizing classical theories in their respective domains, such as the classical control theory, information theory, and optimization theory. As the AVs can make more informed control decisions when equipped with the V2X communication capability for exchanging sensory information with surrounding vehicles and infrastructure, it is increasingly recognized that the performance of autonomous driving can be significantly enhanced through the joint design of control and communication.  Existing efforts in pursuing this holistic approach often adopt either a \textit{control perspective} or \textit{communication perspective}, focusing on designing either  “\textit{an efficient AV control (AVC) approach with awareness of the communication nonideality such as delay and packet loss}” \cite{ma2020distributed} or “\textit{an efficient communication approach taking cognizance of the requirements to support the AVC applications}” \cite{parvini2023aoi}. In our point of view, the co-design of control and communication is loosely-coupled in the above research, since the emphasis is still on the design of individual subsystems.

%Although the loosely-coupled co-design approach optimizes one subsystem with awareness of the other subsystem, the level of cognizance is limited since a simplified model or conventional design of the other subsystem is normally considered. The inaccurate or conservative modelling could result in suboptimal solution. 

Although the loosely-coupled co-design approach optimizes one subsystem with awareness of the other subsystem, the level of cognizance is limited since a simplified model or conventional design of the other subsystem is normally considered. The inaccurate or conservative modeling could result in a significant deviation from the optimal solution. For example, the overwhelming majority of existing Radio Resource Allocation (RRA) algorithms were developed under the assumption of a conventional AV controller with little or no consideration on the communication nonideality. Consequently, they tend to allocate more resources than necessary for transmitting AVC messages if an advanced AV controller that is robust to communication nonideality is used in practice. Therefore, the interplay between the control and communication subsystems underscores the need for a \textit{tightly-coupled} co-design approach.

The core of the \textit{tightly-coupled} co-design approach lies in the requirement that the communication subsystem optimizes "\textit{how the information is disseminated}" with sufficient and accurate knowledge of "\textit{how the information is leveraged}" in the control subsystem, and vice versa. Moreover, the two subsystems should be \textit{optimized simultaneously} so that each subsystem is aware of the optimal design of the other subsystem to calibrate its own design.  As we navigate through the intricacies of this ambitious approach, three fundamental research questions emerge, i.e.,

\begin{itemize}
\item What theoretical tool(s) should be harnessed for the co-design?
\item How to capture the interplay between control and communication in the conceived optimization problems?
\item 	Where should the co-design and joint optimization be performed, given the computation complexity and the knowledge sharing requirements between the two subsystems?
\end{itemize}

This article presents a solution for the tightly-coupled co-design of control and communication, addressing the aforementioned questions.\par 

In response to the first question, we propose to design an integrated Deep Reinforcement Learning (DRL)-based decision-making system, where DRL is employed to optimize AVC and RRA in each subsystem \cite{lei2020deep}. Recent years have witnessed a growing interest in applying data-driven Machine Learning (ML) to autonomous driving as well as vehicular communications. As a combination of deep learning (DL) and reinforcement learning (RL), the DRL approaches have been demonstrated to outperform the traditional model-driven control and optimization approaches, since less restrictive assumptions on the stochastic properties of the system dynamics are required \cite{7792374}. \par  

To implement the tightly-coupled co-design philosophy, we first formulate a multitimescale decision process, wherein a single policy makes control and communication decisions across different time scales to optimize both performance aspects. Given the complexity of solving such a large-scale optimization problem, we decouple the multitimescale decision process into a control MDP and a communication MDP, utilizing DRL algorithms to solve each model. However, since the two MDP models are inherently interrelated, the optimal policy for the control MDP cannot be determined without knowledge of the optimal policy for the communication MDP, and vice versa. To resolve this dilemma, we use an iterative training approach, where both MDP models are sequentially solved in each iteration. \par

To address the second question, we leverage the Value of Information (VoI) for integrating the control and communication subsystems. The concept of VoI was first introduced by Howard in 1966~\cite{VoIHoward}, which generally accounts for how information impacts the outcome of subsequent decisions. However, Howard did not provide a rigorous definition or mathematical formula for calculating VoI. Building on his work, various definitions of VoI have been proposed to suit different applications and objectives, such as KL divergence and information quality~\cite{Alawad2022}. Unlike previous studies, this article introduces a systematic approach to defining and evaluating VoI, rooted in the MDP, optimal control, and RL theories. The proposed approach can be applied to any decision making systems that are designed based on the aforementioned theories. In addition, we use VoI as a crucial link between control and communication design. The VoI is estimated by the control module at the information recipient and provided to the communication module at the information source. The VoI is then used to derive the reward function of communication MDP and guide the optimization of communication decisions. In contrast to other information metrics, such as Age of Information (AoI), our proposed VoI has the distinct advantage of accurately reflecting the impacts of information-induced control performance loss~\cite{VoIControl}.\par

For the third question, we advocate joint optimization performed in the virtual space of the VDTN, fully leveraging its potential of powerful computational capability and massive amount of available data reflecting the physical space. Given the communication nonideality between the physical and virtual spaces, we adopt the Centralized Training Decentralized Execution (CTDE) approach. In this approach, deep neural networks (DNNs) are deployed on physical entities, such as vehicles, to make real-time local decisions in a decentralized manner. Meanwhile, these DNNs are centrally trained in the virtual space, where sensory information from various physical entities for both control and communication subsystems is shared to optimize the models. The trained models are continuously fine-tuned in the virtual space to adapt to non-stationary environment, with model parameters periodically transmitted to the physical space to update the corresponding DNNs. \par

In the following sections, we will delve deeper into each component of the joint optimization framework, further elucidating their intricacies, interrelationships, and respective contributions to the overarching goal of joint optimization of control and communications. Specifically, this paper first presents the system architecture of the VDTN in Section~\ref{sec:VDTN}, laying the groundwork for the subsequent study. Building upon this architecture, we then formally define the VoI in the integrated DRL-based decision making system in Section~\ref{sec:VoI}, with the target of shedding light on the co-design of control and communication. Next, we propose the framework for VoI-driven joint optimization of control and communication in the VDTN in Section~\ref{sec:JointOpti}. This proposed framework is characterized by a detailed description, outlining the steps and methodologies involved in achieving the joint optimization objectives. A case study on vehicular platoon applications in the VDTN is performed in Section~\ref{sec:Results} to validate the effectiveness of the proposed framework. Finally, our conclusions and outlooks are given in Section~\ref{sec:Conclusion}.

   \begin{table*}[htb!]
			\renewcommand{\arraystretch}{1.0}
		\setlength{\extrarowheight}{2.2pt}
		\centering
 		\caption{Summary of important symbols used}
%		\begin{tabular}{c|c|l}
    \begin{tabular}[b]{p{2.4cm}<{\raggedright}p{2.2cm}<{\raggedright}p{11cm}<{\raggedright}}
			\hline
			\textbf{Category} & \textbf{Symbol}&\textbf{Definition} \\
			\hline
            \specialrule{0em}{1pt}{1pt}
			\multirow{12}{*}{\textbf{Control}}
            &$k$& The index of a control interval\\
           \cline{2-3} 
			&$S_k^{\text{CL}}$ & The control state at control interval $k$ \\
           \cline{2-3} 
			&$O_k^{\text{CL}}$ & The impaired observation of control state at control interval $k$\\
            \cline{2-3} 
			&$\widetilde{S}_k^{\text{CL}}$ & The augmented control state at control interval $k$ \\
           \cline{2-3} 
			&$u_k^{\text{CL}}$& The control action at control interval $k$\\
           \cline{2-3}  
			& $R_{k+1}^{\rm CL}$ & The control reward at control interval $k$\\
           \cline{2-3} 
			&$\gamma$& The discount factor\\
           \cline{2-3} 
           &$\widetilde{\pi}_{\rm CL}$& The acting control policy based on impaired observation\\
           \cline{2-3} 
           & $ \pi_{\rm CL}^{*}$ & The optimal control policy based on perfect observation\\
           \cline{2-3} 
           & $\widetilde{\textbf{J}}_{\rm CL}$ & The control performance of the acting control policy $\widetilde{\pi}_{\rm CL}$\\
			\cline{2-3} 
			&$\textbf{J}_{\rm CL}^{*}$& The control performance of the optimal control policy $ \pi_{\rm CL}^{*}$\\
           \cline{2-3}
           &$A_{\pi_{\rm CL}^{*} }(S_k^{\rm CL}, u_k^{\rm CL})$ & The advantage function of the optimal control policy $ \pi_{\rm CL}^{*}$\\
     		 \hline
			\multirow{9}{*}{\textbf{Communication}}
			&$t$ & The index of a communication interval within a control interval\\
           \cline{2-3} 
			&$T$& The total number of communication intervals within a control interval \\
            \cline{2-3}
            &$S^{\rm CM}_{(k,t)}$&The communication state at communication interval $(k,t)$\\
           \cline{2-3} 
           &$u^{\rm CM}_{(k,t)}$&The communication action at communication interval $(k,t)$\\
           \cline{2-3} 
           &$R^{\rm CM}_{(k,t+1)}$&The communication reward at communication interval $(k,t)$\\  
           \cline{2-3} 
            & $ \pi_{\rm CM}$ & The communication policy \\
            \cline{2-3} 
			&$\textbf{J}_{\rm CM}$& The communication performance of the communication policy $ \pi_{\rm CM}$\\
            \cline{2-3} 
            &$\mathrm{P}_{\rm CM}$ & The expected long-term communication performance of traditional services \\
            \cline{2-3} 
            &$g_{\text{CM}}(S_{k,t}^{\text{CM}}, u_{k,t}^{\text{CM}})$ & The instantaneous communication performance of traditional services at communication interval \((k,t)\) \\
           \hline
           \multirow{2}{*}{\textbf{VoI}} 
           &$\Bar{\Xi}$& EVoI \\
           \cline{2-3}
           &$\xi_{k}$ & IVoI at control interval $k$ \\
            \hline
		\end{tabular}
		\label{symbol}
	\end{table*}

\section{System Architecture of the VDTN}
\label{sec:VDTN}

\begin{figure*}[!t]
    \centering
    \includegraphics[width=0.9\linewidth]{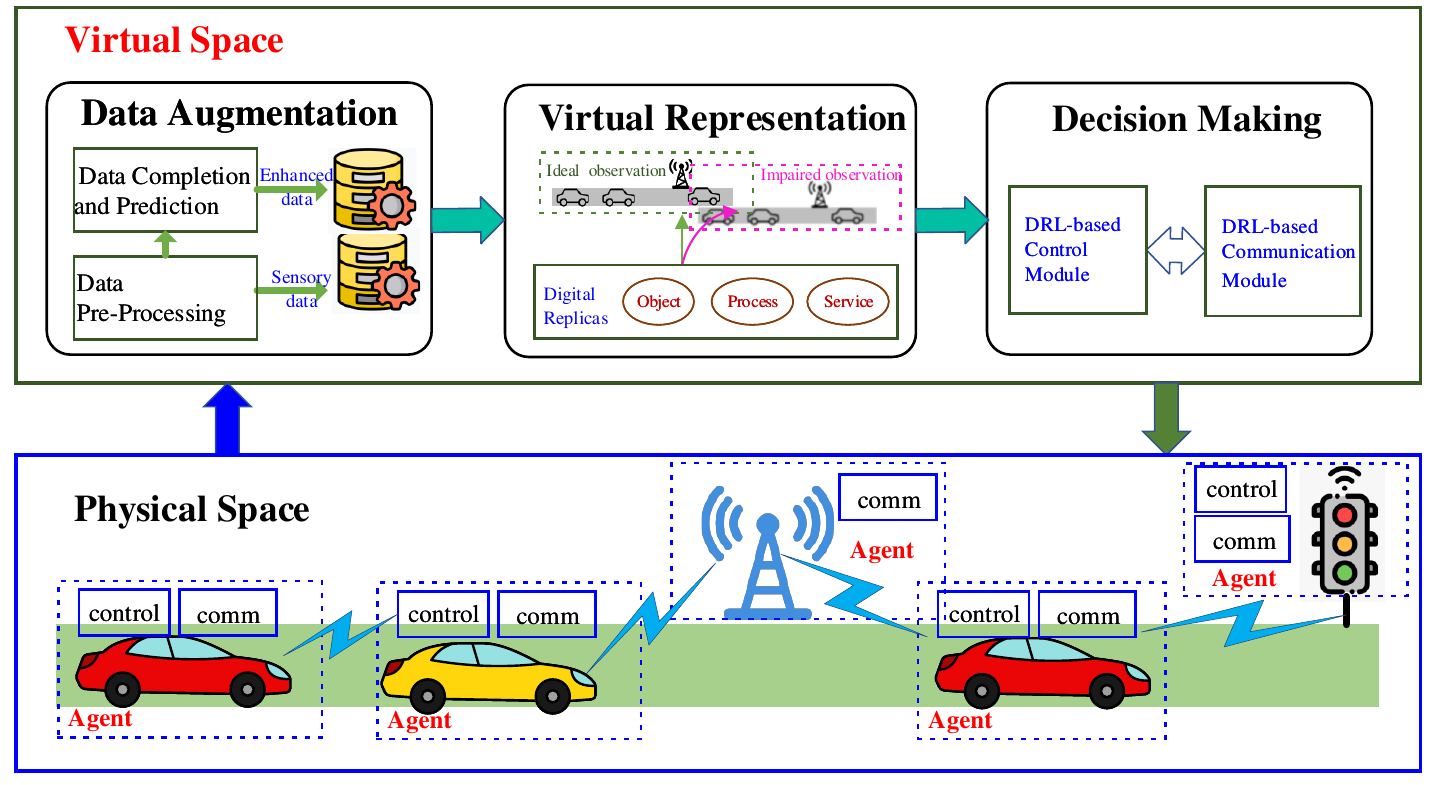}
    \caption{Illustration of the system architecture of the Vehicular Digital Twin Network (VDTN).}
    \label{figVDTN}
\end{figure*}

As shown in Fig.~\ref{figVDTN}, in the VDTN, various entities including vehicles, road infrastructure such as traffic lights, and communication infrastructure such as base stations collect real-time sensory data, which are then transmitted to the edge or cloud computing servers through 6G network to construct the virtual replicas. In the virtual space, a massive amount of data undergoes pre-processing, completion, prediction, and other operations to generate the virtual representation, which is essential for the subsequent decision-making modules. \par

In an VDTN ideally powered by 6G, a huge amount of sensors deployed in the physical space reliably and promptly transmit data to the virtual space. Meanwhile, the decisions made in the virtual space are delivered in real-time and executed by actuators in the physical space. Consequently, tasks traditionally challenging in conventional IoV due to complexity limitations, such as joint optimization of control and communication, can be effectively performed in the VDTN. \par

However, in the early stages of 6G development, limited radio resources and non-ideal transmission performance continue to be constraints for the VDTN. There are two common types of wireless connections in the VDTN, i.e., \textit{those between various entities in the physical space} and \textit{those between the physical and virtual spaces}. In this paper, we assume that both types of connections are non-ideal, meaning that data transmission may experience impairments such as delay and packet loss. This assumption reflects real-world scenarios in the near future. \par

For the sake of illustration, this section provides only a brief description of several DT modules related to AVC and RRA, which are the focus of this paper. However, it does not mean that the VDTN is only limited to these modules and functions.

\subsection{Data Augmentation}

There are mainly two functionalities in this module, i.e.,
\begin{itemize}
\item \textbf{Data Pre-Processing:} This involves cleaning and organizing the extensive sensory data, thoroughly removing inconsistencies, errors, and redundancies to ensure data integrity. The refined data is systematically stored across various databases, encompassing both structured and unstructured formats. This systematic storage approach enhances the efficiency of data retrieval and utilization in the VDTN.
 
\item \textbf{Data Completion and Prediction:}  Owing to factors such as measurement errors or communication nonideality, not only the sensory data exchanged between the entities in the physical space but also those received at the virtual space could be impaired. In the virtual space, sophisticated deep learning models such as Generative Nets and Transformers can be harnessed~\cite{DTGAN}, which exploit historical and current data to complete/predict the missing/delayed data, resulting in the \textit{enhanced data}. Subsequently, this new data can be used to achieve a more realistic representation of the physical space. 
\end{itemize}

\subsection{Virtual Representation}
\label{subsec:virtual}
All objects, processes, and services in the physical space of IoV can be represented in the virtual space using the enhanced data. The data used to create these digital replicas includes a diverse set of information, encompassing both road traffic and vehicular communication characteristics. The former comprises vehicle behaviors, road conditions, and so on. The latter includes the radio channel environment, communication protocols, service characteristics, network performance measures, and so forth. From the perspective of control and communication co-design, the virtual representation in VDTN includes the digital replicas for the control subsystem and the communication subsystem, respectively.\par

From the perspective of RL, the virtual representation provides digital replicas of the environment and agents in the physical space. Despite the non-ideal communication between the physical and virtual spaces, an exact virtual replica of the physical environment could be constructed from the enhanced data as described in Section II.A, which means that the states in the physical and virtual spaces are exactly the same for each time step. Moreover, the enhanced data could also be leveraged to learn a model of the physical environment including transition probability and reward function. Based on the learned model, the virtual space could also generate an environment that evolves in the same stochastic manner as that of the physical space. Training agents in the virtual space enjoys much more advantages than in the physical space. For example, the environment can be replicated into multiple instances based on the learned environment model just as the multiple sample realizations of the same stochastic process, while an agent is split into multiple copies that are executed in parallel to obtain different experiences. \par  

In this paper, we consider that two types of environment are conceived for the control subsystem based on these digital replicas, namely, the control environment with \textit{perfect observation} and the control environment with \textit{impaired observation}. The perfect observation refers to the perceived information about the environment that the agent can obtain under the assumption of ideal transmission in V2X links. This type of control environment can be generated based on the enhanced data as described above. However, in the physical space, due to the non-ideal nature of V2X communication between the physical entities, delay or packet loss may occur during transmission, leading to inaccuracies in the information received by the physical entities. Based on such information, the agent can experience impaired observations of its environment. In the virtual space, we can simulate a control environment with impaired observation by deriving it from the control environment with perfect observation, while incorporating statistical insights from the communication environment regarding non-ideal communication between physical entities. Both types of environment will be used in the joint optimization of control and communication as elaborated in Section~\ref{sec:JointOpti}. \par

\subsection{Decision Making}

Most of IoV services necessitate the simultaneous utilization of communication and control-related information for timely and intelligent decision-making. These decisions span a range of tasks, including intelligent transportation aspects such as AVC, traffic flow control, and traffic route planning, as well as communication-related aspects such as deciding when or whether to share information and RRA. While our focus in this paper revolves around AVC and RRA decisions, the proposed methodology can be extended to address various other control and communication decisions. Following RL conventions, the decision-makers for AVC and RRA are referred to as the control agent and communication agent, respectively.\par

Ideally, decisions are made in the virtual space and transmitted to the physical space for execution in real-time. However, considering the strict real-time requirements of safety-related IoV services, the V2X network has difficulty in promptly and accurately acquiring sensory information from and disseminating decisions to the diverse entities in the physical space. Therefore, adopting the approach of \textit{CTDE} is more practically feasible in the near future. In the physical space, agents are deployed on vehicles and base stations, enabling real-time local control and communication decisions in a decentralized manner. Specifically, each DRL-based agent is usually equipped with a DNN  for generating the local actions, where the DNN is either a local actor representing the per-agent policy function or a local Q-network representing the per-agent action-value functions. In the virtual space, these DNNs can be centrally trained by leveraging the large amount of computing power as well as the massive amount of spatial-temporal data. Once essentially converged, the DNN models are swiftly deployed to agents at base stations and vehicles through V2X links. Given the dynamic nature of IoV environments, including road conditions and communication network dynamics, the trained models in the virtual space require periodic and continuous tuning to adapt to these changes. Updated models should be disseminated to the physical space to ensure the effectiveness of decision-making in response to the dynamic environment of IoV. Since models are updated to adapt to changes in the stochastic properties of the environment, the frequency of model updates is much lower than the frequency of decision-making. Consequently, the 6G network is well-suited for disseminating the updated models to the physical space in a timely manner.  
 
It is noteworthy that CTDE is a widely adopted paradigm for addressing multi-agent DRL problems, where a DNN, namely the centralized critic representing the joint action-value function, is typically trained to assist in training multiple decentralized per-agent DNNs~\cite{multiagent}. In our context, CTDE not only tackles the multi-agent problems for AVC and RRA. More importantly, it is essential for the co-design of AVC and RRA, where the VoI estimated by the control subsystem of information recipient can be shared with the information disseminator to guide the optimization of communication subsystem, as elaborated in the subsequent sections of the paper.

\section{DRL-based VoI for Decision Making}
\label{sec:VoI}

\subsection{Multitimescale Decision Process}

In the VDTN, the control and communication subsystems make decisions at different time scales. For example, the sampling and decision-making intervals for AVC normally range between $0.01$ $\rm s$ to $0.1$ $\rm s$. Meanwhile, the RRA decisions for vehicular communication are made per subframe, e.g., with a duration of $1$ $\rm ms$ in the LTE-V2X systems~\cite{5GRRA}. Therefore, the time steps in the MDP model for AVC and RRA have different lengths, and are therefore referred to as the control intervals and communication intervals, respectively, whenever it is necessary to distinguish between them. We consider that a control interval consists of $T$ communication intervals. In the rest of the paper, the control intervals are indexed by $k \in \{0, 1, 2, \ldots\}$, while the communication intervals in control interval $k$ are indexed by $(k, t)$, where $t \in \{0, 1, 2, \ldots, T-1\}$. \par 

As such, the co-design problem of control and communication can be conceived as a multitimescale decision process, where a single policy prescribes the control and communication actions at different time scales, with the goal of optimizing the control performance as well as the communication performance. However, such a full-state approach yields a large-scale optimization problem, which is computationally infeasible even for the VDTN. \par

To overcome the above challenge, we decouple the single full-state MDP model into two separate MDP models, i.e., a control MDP and a communication MDP, where each MDP model is solved by the DRL algorithms for making the control and communication decisions, respectively. Thus, the integrated decision making system comprises a DRL-based control module and a DRL-based communication module. However, as the environment of the communication MDP is affected by the control policy and vice versa, the two MDP models are interrelated as will be explained in detail below, where the optimal policy of one model depends on that of the other model.  \par

\subsection{Interplay between Control and Communication}

\subsubsection{Impact of the RRA policy on control}

In RL, the environment state includes all the information that the environment uses to determine the next state and reward. However, the environment state is usually not fully and accurately visible to the agent. Therefore, the agent has to choose actions based on its internal representation of the state, which is normally called the observations.

As discussed in Section~\ref{sec:VDTN}, the control agents in VDTN are deployed in the vehicles for making real-time decisions in a decentralized fashion, after they are centrally trained in the virtual space. Thus, the observations of each control agent are formed by the aggregation of the local information sampled by the sensors equipped on its own vehicle, as well as the V2X information sampled by and transmitted from the remote sensors on the other vehicles and the infrastructure. Since the wireless transmission in IoV is usually not ideal, the sensory data via V2X links could be delayed or lost. In such a case, the control agent has to make decisions based on the impaired observations $O_k^{\text{CL}}$ instead of the perfect observations of the state $S_k^{\text{CL}}$. Therefore, the vehicle control problem becomes a partial observable MDP (POMDP).

The POMDP can normally be converted into an MDP using state augmentation, where the augmented control state $\widetilde{S}_k^{\text{CL}}$ includes not only the impaired observation $O_k^{\text{CL}}$, but also the action-observation history as well as some auxiliary information such as the delay of the current observation. The optimization objective of the DRL-based control module can be written as
\begin{align}
\label{VCObjective}
\max_{\widetilde{\pi}_{\rm CL}} \widetilde{\textbf{J}}_{\rm CL} =   \max_{\widetilde{\pi}_{\rm CL}}  \mathbb{E}_{\widetilde{\pi}_{\rm CL},\pi_{\rm CM}} [\sum_{k=0}^{\infty} \gamma^k R_{k+1}^{\rm CL} ],
\end{align}
where $R_{k+1}^{\rm CL}$ is the immediate reward that the control agent receives at control interval $k$, and $\gamma \in [0, 1]$ is the discount factor. The subscript of $\mathbb{E}$ denotes the expected value of a random variable when the control agent follows policy $\widetilde{\pi}_{\rm CL}$ and the communication agent follows policy $\pi_{\text{CM}}$. Equation~\eqref{VCObjective} reveals that the control performance $\widetilde{\textbf{J}}_{\rm CL}$ is not only affected by the control policy $\widetilde{\pi}_{\rm CL}$ but also the RRA policy $\pi_{\text{CM}}$. The impact of $\widetilde{\pi}_{\rm CL}$ on $\widetilde{\textbf{J}}_{\rm CL}$ is obvious, since given the augmented state $\widetilde{S}_k^{\text{CL}}$ at control interval $k$, the control decision on the action $u_k^{\text{CL}}$ is made according to the control policy $\widetilde{\pi}_{\rm CL} (u_k^{\rm CL} |\widetilde{S}_k^{\rm CL})$. Meanwhile, given the observation $\widetilde{S}_k^{\text{CL}}$ and action $u_k^{\text{CL}}$, the next state $\widetilde{S}_{k+1}^{\rm CL}$ is affected by the RRA policy, i.e., $\mathrm{Pr}(\widetilde{S}_{k+1}^{\rm CL}|\widetilde{S}_k^{\rm CL},u_k^{\rm CL},\pi_{\rm CM})$. This is because the RRA policy affects the stochastic properties of communication nonideality, such as delay distribution, which in turn impacts the impaired observation received by the control agent at the next control interval.

Since the RRA policy is an integral part of the control environment, the DT must be aware of the RRA policy or the resulting data from communication nonidealities to emulate a realistic environment when training the control agent.

\subsubsection{Impact of the AVC policy on communication}

The optimization objective of the DRL-based communication module in the VDTN is typically twofold – optimizing network performance metrics for traditional services and minimizing control performance loss due to communication nonidealities for control-related services. For this purpose, the communication agent should obviously allocate the scarce radio resources to carry the more meritorious information with higher priority, so that the information is  subject to less impairment during transmission. To quantify which information is more “meritorious” for the control system, we will formally define the VoI in Section~\ref{subsec:VoI}. \textit{The VoI serves as a bridge between the two MDP models for control and communication, which is estimated by the control agent who receives and acts upon the V2X information; and optimized by the communication agent who transmits the V2X information.} Note that the merit of the information depends on how it is leveraged by its recipient. Therefore, the VoI depends on AVC policy, as the more robust the policies are to the communication nonideality, the less important it is to transmit the information without impairment.

\subsection{VoI Definition}
\label{subsec:VoI}

The term VoI has been studied in different research communities for various decision-making tasks. It is generally used to measure the benefit of acquiring additional information at the time of making decisions. For the purpose of joint control and communication optimization in the VDTN, we use VoI to quantify the performance loss that would result from making AVC decisions based on the impaired V2X information instead of the perfect information, where the impairment is caused by the communication nonideality such as packet delay and loss.

In contrast to traditional communication performance metrics like throughput and delay, along with widely-used control-oriented metrics such as AoI, VoI emerges as a more apt gauge for optimizing the communication system. This suitability stems from its ability to well capture the influence of communication nonidealities on control performance. Next, we define two types of VoI, which are used to formulate the MDP model for the communication systems. 

 \textbf{Expected Cumulative VoI (EVoI):}

The EVoI $\Bar{\Xi}$ is defined as the difference in the performance of the acting control policy $\widetilde{\pi}_{\rm CL}$  based on the impaired observation  and the optimal control policy $ \pi_{\rm CL}^{*}$ based on the perfect observation, respectively, i.e., $ \Bar{\Xi}=\widetilde{\textbf{J}}_{\rm CL} -\textbf{J}_{\rm CL}^{*}$.

The EVoI indicates the performance loss as a consequence of the control agent consistently receiving the impaired observations at every control interval. In the terminology of RL, EVoI corresponds to the regret of the acting control policy $\widetilde{\pi}_{\rm CL}$. It is worth noting that the EVoI is a non-positive value, with a larger EVoI indicating less performance loss. When the EVoI reaches its maximum value of zero, no performance loss is experienced.
%, and the performance of $\widetilde{\pi}_{\rm CL}$ based on impaired observations matches that of $ \pi_{\rm CL}^{*}$ based on perfect observations.

 \textbf{Immediate VoI (IVoI):}

Given the perfect observation $S_k^{\rm CL}$ and impaired observation $O_k^{\rm CL}$ at control interval $k$, the IVoI $\xi_k$ is defined as the difference in the performance starting from control interval $k$, where the control decision at control interval $k$ is made by the acting control policy $\widetilde{\pi}_{\rm CL}$ based on the impaired observation and the optimal control policy $ \pi_{\rm CL}^{*}$ based on the perfect observation, respectively, while the rest of the sequential control decisions are made by $ \pi_{\rm CL}^{*}$ based on the perfect observations $S_{k+1}^{\rm CL},S_{k+2}^{\rm CL},\dots$, i.e., $\xi_k=A_{\pi_{\rm CL}^{*} }(S_k^{\rm CL}, u_k^{\rm CL})|_{u_k^{\rm CL}\sim\widetilde{\pi}_{\rm CL}(\cdot|\widetilde{S}^{\rm CL}_k)}$, where $u_k^{\rm CL}\sim\widetilde{\pi}_{\rm CL}(\cdot|\widetilde{S}^{\rm CL}_k)$ indicates that the action $u_k^{\rm CL}$ is sampled according to $\widetilde{\pi}_{\rm CL}$.

The IVoI indicates the performance loss as a consequence of the control agent receiving the impaired observations only at a specific control interval $k$, while consistently obtaining perfect observations throughout the remaining time. Moreover, IVoI is calculated for a specific state and observation pair at control interval $k$. In other words, IVoI reveals that if the perfect observation at control interval $k$ is $S_k^{\rm CL}=s$, what is the performance loss if the agent only has access to the impaired observation $O_k^{\rm CL}$ instead? In the terminology of RL, IVoI corresponds to the advantage function of the optimal policy $\pi_{\text{CL}}^*$. Similar to EVoI, IVoI is also a non-positive value, where maximizing IVoI leads to minimizing the performance loss.

It is proved in \cite{VoILei} that the EVoI is equivalent to the expected cumulative IVoI, similar to the relationship between the expected cumulative reward and immediate reward in RL. As a result, we name the two types of VoI as the expected cumulative VoI and immediate VoI, respectively. However, it is essential to clarify that the term immediate VoI doesn't imply its sole focus on quantifying the disparity in the immediate rewards at control interval $k$ when the control decisions are made based on the impaired versus perfect observations, respectively. Since the decision at control interval $k$ not only affects the immediate reward at the current control interval but also the future rewards through its impact on the next state, the IVoI measures the performance loss in terms of the expected cumulative reward just as the EVoI does.

\section{Joint Optimization of Control and Communication}
\label{sec:JointOpti}

\begin{figure*}[!t]
    \centering
    \includegraphics[width=0.8\linewidth]{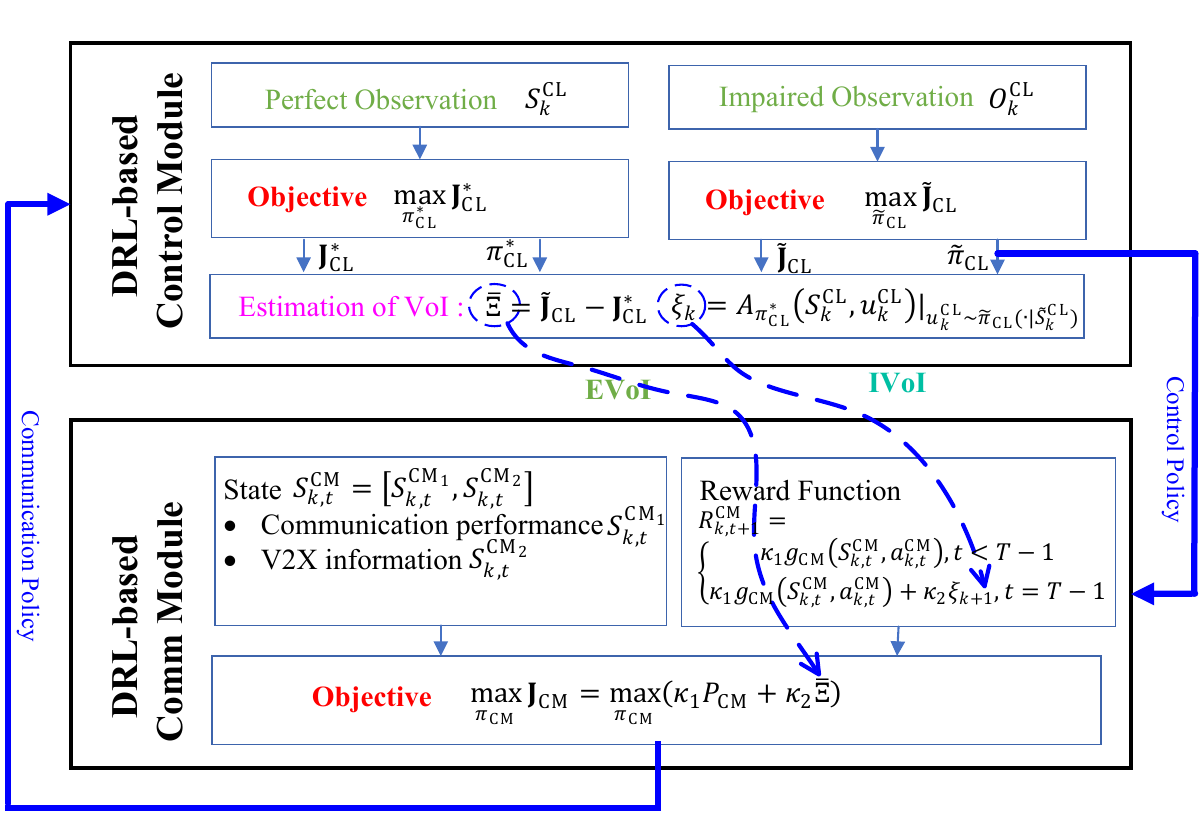}
    \caption{Framework of Joint Optimization of Control and Communication in  VDTN.}
    \label{figFramework}
\end{figure*}

In general, the objective is to maximize communication performance as well as the control performance through the optimization of the AVC and RRA policy in the VDTN. As discussed in Section IV, we can conceive two MDP models for the control subsystem and communication subsystem, respectively, utilizing DRL algorithms for their solutions. \par

Due to the inherent interplay between control and communication, the optimal policy of one MDP model cannot be derived without knowing the optimal policy of the other model. To address this dilemma, an iterative training approach is adopted to find the optimal strategy of the control MDP and communication MDP simultaneously \cite{VoILei}. In each iteration, both agents are sequentially trained, and their optimal policies under the current environments are obtained. Notably, the control agent is consistently trained before the communication agent, facilitating the update of the VoI and providing it to the communication agent for further optimization of communication policy. In the initial iteration, a random communication policy can be employed for training the control agent. \par 

The iterative joint optimization framework is illustrated in Fig. 2. In this framework, the control policy derived by the DRL-based control module is provided as input to the DRL-based communication module. Similarly, the communication policy derived by the DRL-based communication module is provided as input to the DRL-based control module. In the following, we delve into further details on how the communication and control agents are trained to learn their respective optimal policies in each iteration. \par

\subsection{Learning of communication-aware control policy}
From the above discussion, the DRL-based control module needs to be communication-aware, which means the control policy is learned under the assumption that a communication policy is given. Note that the communication policy is an integrated part of the control environment, since it affects communication nonideality statistics, which in turn influences the impaired observations at the control agent. As discussed in Section IV.B, an MDP model for the AVC problem with the optimization objective as shown in \eqref{VCObjective} can be formulated. Subsequently, state-of-the-art DRL algorithms can be used to acquire the desired control policy. Additionally, to further engage in joint optimization with the communication subsystem, it is necessary to obtain the VoI associated with the control performance.  \par

To obtain either EVoI or IVoI, the DRL-based control module has to train the optimal control policy $\pi_{\text{CL}}^*$ in addition to the acting control policy $\widetilde{\pi}_{\text{CL}}$. As discussed in Section~\ref{subsec:virtual}, two types of environment are created in the virtual representation. As shown in Fig. 2, the acting control policy $\widetilde{\pi}_{\text{CL}}$ is learned in the environment with impaired observations, while the optimal control policy $\pi_{\text{CL}}^*$ is learned in the environment with perfect observations. After training the two control policies, the EVoI can be derived by simply calculating the performance difference between two policies. \par

The IVoI corresponds to the advantage function of the optimal control policy with perfect observations, where different methods can be used to estimate the advantage function of a policy in RL, such as by the Temporal Difference (TD) error $\delta_{\widetilde{\pi}_{\text{CL}},k}= R_{k+1}^{\text{CL}} + v_{\pi_{\text{CL}}^*}(S_{k+1}^{\text{CL}}) - v_{\pi_{\text{CL}}^*}(S_k^{\text{CL}})$. In this case, the IVoI can be derived for each control interval based on the reward $R_{k+1}^{\text{CL}}$ as a result of implementing action $u_k^{\text{CL}}\sim\widetilde{\pi}_{\text{CL}}(\cdot|\widetilde{S}_k^{\text{CL}})$, as well as the value function $v_{\pi_{\text{CL}}^*}$ of the optimal control policy for the current state $S_k^{\text{CL}}$ and the next state $S_{k+1}^{\text{CL}}$. \par

\subsection{Learning of control-aware RRA policy:}
Similarly, the DRL-based communication module has to be control-aware, implying that the communication policy is learned under the assumption that a control policy is given. Specifically, the VoI is estimated within the control module of the information recipient and is subsequently used to guide the optimization within the communication module of the information disseminator.

As discussed in Section IV.B, the objective of the DRL-based communication module is to optimize the traditional network performance metrics as well as to minimize the control performance loss due to communication nonideality, which is captured by the EVoI as defined in Section~\ref{subsec:VoI}. As a result, the optimization objective is given as   
\begin{align}
\label{RRAObjective}
\max_{\pi_{\rm CM}} \textbf{J}_{\rm CM} =   \max_{\pi_{\rm CM}} 
( \kappa_1 \mathrm{P}_{\rm CM} + \kappa_2 \Bar{\Xi}),
\end{align}
where \(\mathrm{P}_{\rm CM}\) is the expected long-term communication performance of traditional services, such as the expected cumulative throughput over the considered time horizon, and \(\Bar{\Xi}\) is the EVoI. The weighting coefficients \(\kappa_1\) and \(\kappa_2\) reflect the relative importance of traditional network services and control services in the optimization objective. They can be conveniently set according to control performance requirements, availability of radio resources, and other factors, satisfying the constraints \(0 \leq \kappa_1, \kappa_2 \leq 1\) and \(\kappa_1 + \kappa_2 = 1\).

Correspondingly, the reward function of the MDP model for the communication subsystem can be designed as
\begin{align}
\label{RRAReward}
R_{k,t+1}^{\text{CM}} =  \begin{cases}
\kappa_1 g_{\text{CM}}(S_{k,t}^{\text{CM}}, u_{k,t}^{\text{CM}}) & \text{if } t < T-1\\
\kappa_1 g_{\text{CM}}(S_{k,t}^{\text{CM}}, u_{k,t}^{\text{CM}}) + \kappa_2 \xi_{k+1} & \text{if } t = T-1
\end{cases}
\end{align}
where \(g_{\text{CM}}(S_{k,t}^{\text{CM}}, u_{k,t}^{\text{CM}})\) represents the instantaneous communication performance of traditional services at communication interval \((k,t)\), such as the instantaneous data rate. The \(g_{\text{CM}}(S_{k,t}^{\text{CM}}, u_{k,t}^{\text{CM}})\) component is present in the reward of each communication interval, while the IVoI component, i.e., \(\xi_{k+1}\), is only present in the last communication interval of each control interval when \(t = T-1\). Note that \(\xi_{k+1} = A_{\pi_{\text{CL}}^*}(S_{k+1}^{\text{CL}},u_{k+1}^{\text{CL}})|_{u_{k+1}^{\text{CL}}\sim\widetilde{\pi}_{\text{CL}}(\cdot|\widetilde{S}_{k+1}^{\text{CL}})}\) represents the amount of performance loss that the information recipient will suffer from, as a consequence of its control agent having to make a decision at the next control interval \(k+1\) based on the impaired observation \(\widetilde{S}_{k+1}^{\text{CL}}\) according to its acting policy \(\widetilde{\pi}_{\text{CL}}\) instead of based on the perfect observation \(S_{k+1}^{\text{CL}}\) according to its optimal policy \(\pi_{\text{CL}}^*\). The impaired observation received at control interval \(k+1\) is affected by the degree of communication nonideality it has experienced, which is in part a result of the sequence of communication actions \(u_{k,0}^{\text{CM}}, u_{k,1}^{\text{CM}}, \ldots, u_{k,T-1}^{\text{CM}}\) that are taken during control interval \(k\). Since EVoI equals the expected cumulative IVoI, the expected discounted sum of the immediate reward in \eqref{RRAReward} equals the optimization objective.\par
As illustrated in Fig. 2, the EVoI and IVoI are estimated by the DRL-based control module and provided to the DRL-based communication module to formulate the communication MDP. \par
The communication state \(S_{k,t}^{\text{CM}}\) consists of two components, i.e., \(S_{k,t}^{\text{CM}} = [S_{k,t}^{\text{CM}_1}, S_{k,t}^{\text{CM}_2}]\). \(S_{k,t}^{\text{CM}_1}\) is related to \(g_{\text{CM}}\), including  Channel State Information (CSI), Queue State Information (QSI), etc., which are typically defined in the MDP models for RRA. Meanwhile, \(S_{k,t}^{\text{CM}_2}\) is related to IVoI, which theoretically should include any information that is useful in predicting the IVoI at the information recipient, such as the V2X information that is to be transmitted. Therefore, the formulation for \(S_{k,t}^{\text{CM}_2}\) differs depending on the specific AVC tasks.

Finally, the DRL algorithms can be employed to determine the optimal communication policy for the DRL-based communication module in each iteration.

\begin{table}[!t]
		\centering
 		\caption{Main parameters of a platoon system supported by VDTN}
    \begin{tabular}[b]{p{5cm}<{\raggedright}p{2cm}<{\raggedright}}
			\hline
   Parameter Description & Value \\
   \hline
\specialrule{0em}{1pt}{1pt}
    Number of vehicles in the platoon & 4 \\
    Number of V2I links & 4 \\
    Control interval of vehicle in the platoon & 50 ms\\
    Size of CAM packet & 1800 Bytes \\ 
    Number of communication intervals within each control interval & 50 \\
    Carrier frequency & 2 GHz \\
    Bandwidth of sub-channel & 180 kHz \\
    Maximum transmit power of vehicle & 23 dBm \\
 \hline 
\end{tabular}
\label{mainparameters}
\end{table}

%%%     AWGN noise power & -90~\myunit{dBm}  

\begin{figure}[!t]
    \centering
    \includegraphics[width=0.9\linewidth]{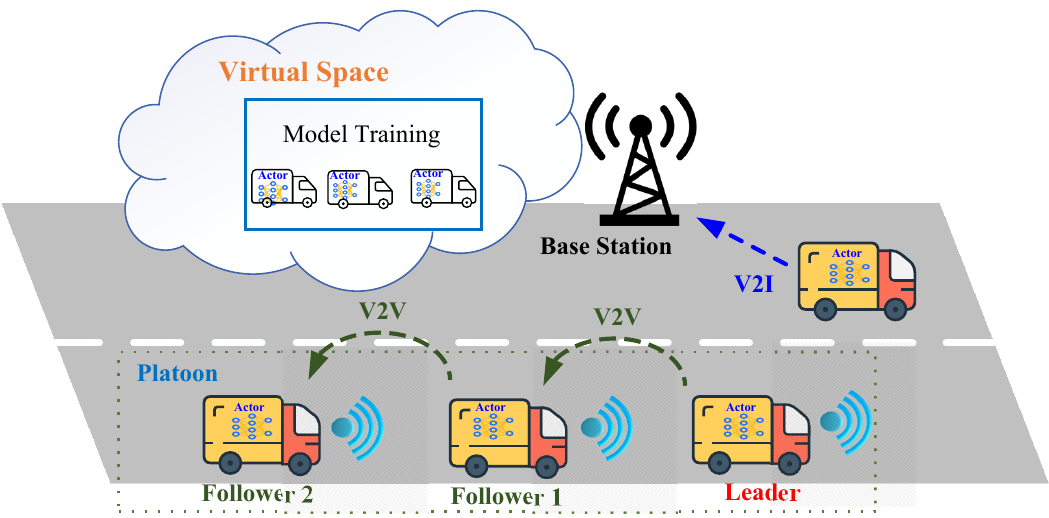}
    \caption{Simulation scenario of the vehicular platoon supported by VDTN.}
    \label{figplatoon}
\end{figure}

%\section{Experimental Results and Analysis}
\section{A Case Study on Vehicular Platoon Applications in the VDTN}
\label{sec:Results}

%\subsection{Application Scenario}

%In the platoon, each control interval is further divided into $T$ communication intervals, i.e., $T=50$ in this case study. Each communication interval has a length of 1 $ms$ corresponding to the subframe duration in C-V2X communication.

In this section, we assess the effectiveness of the proposed joint optimization of control and communication (JOCC) by conducting simulations in a vehicular platoon supported by the VDTN. As shown in Fig. \ref{figplatoon}, the AVs in the platoon communicate using V2V links, while V2I links establish connections between the AVs and the Base Station (BS) for high-throughput services. The Predecessors Following (PF) information typology (IFT) is adopted in the platoon, where the Collaborative Adaptive Message (CAM) of the preceding vehicle is transmitted to its following vehicle. The AVC decisions are made at a control interval of $50$ $\rm ms$, while the RRA decisions are executed at a communication interval of $1$ $\rm ms$.  For comparison purpose, the performance of the Delay-aware RRA (i.e., DRRA) method is also given. DRRA is an improved DRL-based RRA algorithm only aiming to minimize the latency and maximize the success probability of delivering the CAM generated at the beginning of the control interval. Both DRL algorithms are trained/tested where the velocity profile of leading vehicle is obtained from the open-source driving data \footnote{NGSIM.}. The main parameters of the platoon system are given in Table \ref{mainparameters}.  \par

\begin{figure}[!t]
    \centering
    \includegraphics[width=0.45\textwidth]{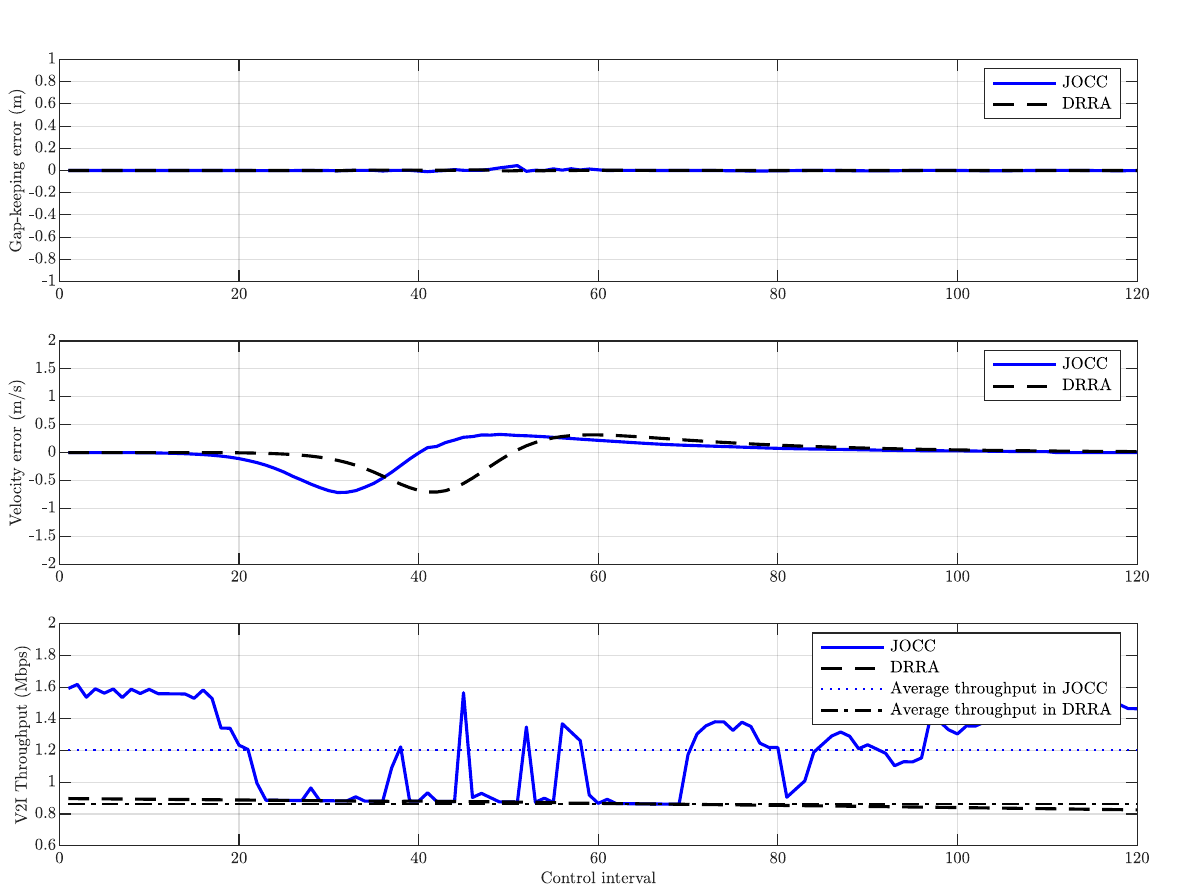}
    \caption{Performance comparison of AVs in the platoon system supported by VDTN.}
    \label{fig_results}
\end{figure}

As illustrated in Fig.~\ref{fig_results}, the control performance of JOCC is minimally impacted in terms of gap-keeping error and velocity error when compared to those of the DRRA. On the other hand, JOCC exhibits a higher average throughput on each V2I link compared to DRRA, achieving $0.86$ $\rm Mbps$ compared to $0.91$ $\rm Mbps$, respectively. This indicates the effective performance of the proposed JOCC in the case of platooning in VDTN.

\section{Conclusions and Outlooks}
\label{sec:Conclusion}

This paper has proposed a VDTN architecture with designed key functional modules for joint control and communication optimization. Aiming to fully utilize the computing capabilities and massive data of the virtual space in digital twin as well as satisfying the real-time requirements, the CTDE framework is adopted in VDTN. Then, we have established a multitimescale decision process for joint optimization of control and communication, incorporating VoI to reflect control performance degradation due to communication nonideality. Furthermore, a joint optimization framework based on DRL has been proposed to address AVC and RRA simultaneously. Finally, simulation results in a platoon scenario have validated the effectiveness of the proposed framework. \par

The anticipated application scope of the proposed VDTN architecture, coupled with the VoI-driven joint optimization approach, is extensive, encompassing a spectrum of vehicle control and communication tasks. The former includes tasks like car following, lane change, overtaking, ramp merging, and negotiating intersections. Concurrently, the latter involves considerations of what to communicate, when to communicate, and how to communicate. In the context of `how-to-communicate', the RRA function discussed in this article plays a pivotal role. The exploration of VoI estimation and VoI-driven optimization in VDTN presents compelling avenues for future research. Some potential research directions include:
\begin{itemize}
\item VoI definition: The VoI definitions in this article can be expanded. For example, VoI can be defined to quantify control performance loss based on whether information is present or absent, rather than whether it is impaired or not.  
\item VoI estimation in the control subsystem: Accurate evaluation of VoI is crucial for guiding the optimization of communication decisions. However, existing RL methods for estimating the advantage function encounter challenges when applied to IVoI evaluation. For instance, the TD error, which serves as a sample of the advantage function, can be noisy. Thus, improving the accuracy of VoI estimation remains an important open issue. 
\item VoI-driven optimization in the communication subsystem: Diverse communication tasks are performed at various time scales, and could, therefore, leverage either the EVoI and/or IVoI as appropriate. Effectively utilizing VoI for optimizing different communication functions is an interesting area for future exploration. 
\end{itemize}

Additionally, several significant challenges must be addressed in the DRL, such as 
\begin{itemize}
	\item multi-agent RL (MARL) issues: Solving both the control and communication MDPs relies on MARL algorithms. However, state-of-the-art MARL algorithms like QMIX and Multiagent Proximal Policy Optimization (MAPPO) are not fully effective in addressing well-known challenges such as non-stationary environment, partial observability, and credit assignment problems. Moreover, the popular CTDE paradigm, used by many MARL algorithms, faces scalability issue attributed to centralized training. Finally, addressing the open ad-hoc team problem is a promising direction for the highly dynamic and non-stationary vehicular environment, where an agent can collaborate with a variable number of other agents without prior coordination.    
    \item long-horizon sparse-reward problem : The time horizon of the communication MDP spans multiple control intervals, with each control interval further divided into multiple communication intervals. Additionally, the VoI reward is solely accessible during the last communication interval of each control interval in \eqref{RRAReward}. This long-horizon, sparse-reward environment presents significant challenges for designing RL algorithms, as highly specific action sequences must be executed prior to observing any nontrivial feedback.    
\end{itemize}

Finally, VDTN faces several deployment issues in real-world scenarios. The first challenge to deploy VDTN is determining the optimal placement of virtual entities, data, and models within the DT domain — specifically, whether to position them on cloud servers or edge servers. For instance, ones that require extensive data processing and computational power are generally better hosted in the cloud, while those with strict real-time requirements are ideally deployed on the edge. Furthermore, the transmission of data and models through various V2X links is another challenge, necessitating an effective strategy for allocating wireless resources to these links.

\bibliographystyle{IEEEtran}
%% Argument is your BibTeX string definitions and bibliography database(s).
\bibliography{IEEEabrv,VDTN}

\begin{thebibliography}{10}
\providecommand{\url}[1]{#1}
\csname url@samestyle\endcsname
\providecommand{\newblock}{\relax}
\providecommand{\bibinfo}[2]{#2}
\providecommand{\BIBentrySTDinterwordspacing}{\spaceskip=0pt\relax}
\providecommand{\BIBentryALTinterwordstretchfactor}{4}
\providecommand{\BIBentryALTinterwordspacing}{\spaceskip=\fontdimen2\font plus
\BIBentryALTinterwordstretchfactor\fontdimen3\font minus \fontdimen4\font\relax}
\providecommand{\BIBforeignlanguage}[2]{{%
\expandafter\ifx\csname l@#1\endcsname\relax
\typeout{** WARNING: IEEEtran.bst: No hyphenation pattern has been}%
\typeout{** loaded for the language `#1'. Using the pattern for}%
\typeout{** the default language instead.}%
\else
\language=\csname l@#1\endcsname
\fi
#2}}
\providecommand{\BIBdecl}{\relax}
\BIBdecl

\bibitem{5GV2X}
H.~Bagheri, M.~Noor-A-Rahim, Z.~Liu, H.~Lee, D.~Pesch, K.~Moessner, and P.~Xiao, ``5{G} {NR-V2X}: Toward connected and cooperative autonomous driving,'' \emph{IEEE Communications Standards Magazine}, vol.~5, no.~1, pp. 48--54, 2021.

\bibitem{6GIoV}
H.~Li, K.~Ota, and M.~Dong, ``Learning {IoV} in 6{G}: {I}ntelligent edge computing for internet of vehicles in 6g wireless communications,'' \emph{IEEE Wireless Communications}, vol.~30, no.~6, pp. 96--101, 2023.

\bibitem{6GDT}
X.~Shen, J.~Gao, W.~Wu, M.~Li, C.~Zhou, and W.~Zhuang, ``Holistic network virtualization and pervasive network intelligence for 6g,'' \emph{IEEE Communications Surveys \& Tutorials}, vol.~24, no.~1, pp. 1--30, 2022.

\bibitem{6GDT2}
X.~Lin, L.~Kundu, C.~Dick, E.~Obiodu, T.~Mostak, and M.~Flaxman, ``6{G} digital twin networks: {F}rom theory to practice,'' \emph{IEEE Communications Magazine}, vol.~61, no.~11, pp. 72--78, 2023.

\bibitem{ma2020distributed}
F.~Ma, J.~Wang, S.~Zhu, S.~Y. Gelbal, Y.~Yang, B.~Aksun-Guvenc, and L.~Guvenc, ``Distributed control of cooperative vehicular platoon with nonideal communication condition,'' \emph{IEEE Transactions on Vehicular Technology}, vol.~69, no.~8, pp. 8207--8220, 2020.

\bibitem{parvini2023aoi}
M.~Parvini, M.~R. Javan, N.~Mokari, B.~Abbasi, and E.~A. Jorswieck, ``{AoI}-aware resource allocation for platoon-based {C-V2X} networks via multi-agent multi-task reinforcement learning,'' \emph{IEEE Transactions on Vehicular Technology}, 2023.

\bibitem{lei2020deep}
L.~Lei, Y.~Tan, K.~Zheng, S.~Liu, K.~Zhang, and X.~Shen, ``Deep reinforcement learning for autonomous internet of things: Model, applications and challenges,'' \emph{IEEE Communications Surveys \& Tutorials}, vol.~22, no.~3, pp. 1722--1760, 2020.

\bibitem{7792374}
C.~Jiang, H.~Zhang, Y.~Ren, Z.~Han, K.-C. Chen, and L.~Hanzo, ``Machine learning paradigms for next-generation wireless networks,'' \emph{IEEE Wireless Communications}, vol.~24, no.~2, pp. 98--105, 2017.

\bibitem{VoIHoward}
R.~A. Howard, ``Information value theory,'' \emph{IEEE Transactions on Systems Science and Cybernetics}, vol.~2, no.~1, pp. 22--26, 1966.

\bibitem{Alawad2022}
F.~Alawad and F.~A. Kraemer, ``Value of information in wireless sensor network applications and the iot: A review,'' \emph{IEEE Sensors Journal}, vol.~22, no.~10, pp. 9228--9245, 2022.

\bibitem{VoIControl}
T.~Soleymani, J.~S. Baras, and S.~Hirche, ``Value of information in feedback control: {Q}uantification,'' \emph{IEEE Transactions on Automatic Control}, vol.~67, no.~7, pp. 3730--3737, 2022.

\bibitem{DTGAN}
B.~Erman and C.~D. Martino, ``Generative network performance prediction with network digital twin,'' \emph{IEEE Network}, vol.~37, no.~2, pp. 286--292, 2023.

\bibitem{multiagent}
R.~Lowe, Y.~Wu, A.~Tamar, J.~Harb, P.~Abbeel, and I.~Mordatch, ``Multi-agent actor-critic for mixed cooperative-competitive environments,'' in \emph{Proceedings of the 31st International Conference on Neural Information Processing Systems}, 2017, p. 6382–6393.

\bibitem{5GRRA}
K.~Sehla, T.~M.~T. Nguyen, G.~Pujolle, and P.~B. Velloso, ``Resource allocation modes in {C-V2X}: From {LTE-V2X} to {5G-V2X},'' \emph{IEEE Internet of Things Journal}, vol.~9, no.~11, pp. 8291--8314, 2022.

\bibitem{VoILei}
L.~Lei, T.~Liu, K.~Zheng, and X.~Shen, ``Multi-timescale control and communications with deep reinforcement learning---{P}art {II}: {C}ontrol-aware radio resource allocation,'' \emph{IEEE Internet of Things Journal}, pp. 1--1, 2024.

\end{thebibliography}

\end{document}